\documentclass[twocolumn,showpacs,superscriptaddress,amsmath,amssymb,nofootinbib,aps, prd]{revtex4-2}
\usepackage{graphicx}
\usepackage{epsfig}
\usepackage{dcolumn}
\usepackage{multirow}
\usepackage{booktabs}
\usepackage{appendix}
\usepackage{makecell}
\usepackage{bm}
\usepackage{amsmath}
\usepackage{overpic}
\usepackage{comment}
\usepackage{color}
\usepackage{subfigure}
\usepackage{threeparttable}
\usepackage{natbib}
\usepackage{lineno}
\usepackage{siunitx}
\usepackage[colorlinks,linkcolor=blue,anchorcolor=blue,citecolor=blue]{hyperref}

\def \gevc {\mbox{GeV/$c$}}
\def \gevcc{\mbox{GeV/$c^2$}}

\begin{document}
\setlength\linenumbersep{7pt}

\title{\bf\boldmath Search for $J/\psi\rightarrow K^{0}_{S}K^{0}_{S}$ and $\psi(3686)\rightarrow K^{0}_{S}K^{0}_{S}$}
\author{\small
M.~Ablikim$^{1}$, M.~N.~Achasov$^{4,c}$, P.~Adlarson$^{76}$, X.~C.~Ai$^{81}$, R.~Aliberti$^{35}$, A.~Amoroso$^{75A,75C}$, Q.~An$^{72,58,a}$, Y.~Bai$^{57}$, O.~Bakina$^{36}$, Y.~Ban$^{46,h}$
% ...
\\
\vspace{0.2cm}
(BESIII Collaboration)\\
\vspace{0.2cm} {\it
$^{1}$ Institute of High Energy Physics, Beijing 100049, People's Republic of China\\
% ...
$^{4}$ Budker Institute of Nuclear Physics SB RAS (BINP), Novosibirsk 630090, Russia\\
% ...
$^{35}$ Johannes Gutenberg University of Mainz, Johann-Joachim-Becher-Weg 45, D-55099 Mainz, Germany\\
$^{36}$ Joint Institute for Nuclear Research, 141980 Dubna, Moscow region, Russia\\
% ...
$^{46}$ Peking University, Beijing 100871, People's Republic of China\\
% ...
$^{57}$ Southeast University, Nanjing 211100, People's Republic of China\\
$^{58}$ State Key Laboratory of Particle Detection and Electronics, Beijing 100049, Hefei 230026, People's Republic of China\\
% ...
$^{72}$ University of Science and Technology of China, Hefei 230026, People's Republic of China\\
% ...
$^{75}$ University of Turin and INFN, (A)University of Turin, I-10125, Turin, Italy; (B)University of Eastern Piedmont, I-15121, Alessandria, Italy; (C)INFN, I-10125, Turin, Italy\\
$^{76}$ Uppsala University, Box 516, SE-75120 Uppsala, Sweden\\
% ...
$^{81}$ Zhengzhou University, Zhengzhou 450001, People's Republic of China\\
\vspace{0.2cm}
$^{a}$ Deceased\\
% ...
$^{c}$ Also at the Novosibirsk State University, Novosibirsk, 630090, Russia\\
% ...
$^{h}$ Also at State Key Laboratory of Nuclear Physics and Technology, Peking University, Beijing 100871, People's Republic of China\\
% ...
}\vspace{0.4cm}}


\begin{abstract}
Using data samples of $(10087\pm 44)\times10^{6}$ $J/\psi$ events and $(2712.4\pm 14.3)\times10^{6}$ $\psi(3686)$ events collected with the BESIII detector at the BEPCII collider, we search for the \textit{CP} violating decays $J/\psi\rightarrow K^{0}_{S}K^{0}_{S}$ and $\psi(3686)\rightarrow K^{0}_{S}K^{0}_{S}$. No significant signals are observed over the expected background yields. The upper limits on their branching fractions are set as \mbox{$\mathcal{B}(J/\psi\rightarrow K^{0}_{S}K^{0}_{S}) <4.7\times 10^{-9}$} and \mbox{$\mathcal{B}(\psi(3686)\rightarrow K^{0}_{S}K^{0}_{S}) <1.1\times 10^{-8}$} at the 90\% confidence level. These results improve the previous limits by a factor of three for $J/\psi\rightarrow K^{0}_{S} K^{0}_{S}$ and two orders of magnitude for $\psi(3686)\rightarrow K^{0}_{S} K^{0}_{S}$.
\end{abstract}

\maketitle

\section{\boldmath Introduction}

Experimental studies of the decays of vector charmonium states $\psi$ [$\psi=J/\psi$ or $\psi(3686)$] to final states consisting of identical bosons, such as $K^{0}_{S} K^{0}_{S}$, serve to test various symmetry principles and conservation laws in the Standard Model. First, studies of these decays are helpful to test Bose symmetry~\cite{Bose_1924}, which implies that a state composed of identical bosons remains unchanged under exchange of any two particles. The $K^{0}_{S} K^{0}_{S}$ final state is forbidden in the $\psi$ decays by Bose-Einstein statistics. Studies of these decays therefore provide valuable data to validate Bose symmetry. Second, investigations of these decays are valuable to search for potential violation of \textit{CP} symmetry beyond the one present in neutral kaon oscillations. The branching fractions of the possible \textit{CP} violating decays $J/\psi\rightarrow K^{0}_{S} K^{0}_{S}$ and $\psi(3686)\rightarrow K^{0}_{S} K^{0}_{S}$, arising from $K^{0}-\Bar{K}^{0}$ oscillations, are predicted to be $(1.94 \pm 0.20) \times 10^{-9}$ for $J/\psi$ and $(0.56 \pm 0.08) \times 10^{-9}$ for $\psi(3686)$ \cite{CPV_psip_2006,CPV_psip_2009}. Third, studies of these decays are also important to test the \textit{CPT} symmetry, which is a general feature of Lorentz-invariant local quantum field theories with a hermitian Hamiltonian. However, some quantum gravity models imply a violation of \textit{CPT} symmetry. In the $K^{0}_{S} K^{0}_{S}$ decay channel, this violation may be quantified with a complex parameter $\omega\sim10^{-3}$~\cite{CPT1_Bernabeu_2004,CPT2_Bernabeu_2006,CPT3_Bernabeu_2006}. This parameter can be constrained using the ratio of the branching fractions of $J/\psi$ or $\psi(3686)$ into $K^{0}_{S} K^{0}_{S}$ and $K^{0}_{S} K^{0}_{L}$. 

 Additionally, the $K^{0}_{S} K^{0}_{S}$ states from $\psi$ decays are sensitive to test quantum nonlocality versus Einstein–Podolsky-Rosen (EPR) locality~\cite{EPR}.  EPR locality within a two-state system of massive particles allows a nonzero yield of $K^{0}_{S} K^{0}_{S}$ states from the spacelike separated coherent quantum system of neutral kaons. The corresponding branching fractions of $J/\psi\rightarrow K^{0}_{S} K^{0}_{S}$ and $\psi(3686)\rightarrow K^{0}_{S} K^{0}_{S}$ are predicted to be $(5.5\pm1.0)\times 10^{-6}$ and $(2.1\pm0.3)\times 10^{-6}$~\cite{EPR_KsKs_2021}, respectively, after considering potential \textit{CP} violation and kaon regeneration. Quantum nonlocality, on the other hand, completely forbids these decays. Searching for these decays can thus provide experimental data for testing the EPR argument.

Experimentally, the MARKIII Collaboration reported the first search for $J/\psi\rightarrow K^{0}_{S} K^{0}_{S}$ by using 2.7 million $J/\psi$ events in 1985, and set an upper limit to its branching fraction at $5.2\times 10^{-6}$ at the 90\% confidence level (CL)~\cite{MARK-III_1985}. The BES experiment searched for $J/\psi\rightarrow K^{0}_{S} K^{0}_{S}$ and $\psi(3686)\rightarrow K^{0}_{S} K^{0}_{S}$ with 58 million $J/\psi$ events and 14 million $\psi(3686)$ events, yielding upper limits of $1.0\times 10^{-6}$ and $4.6\times 10^{-6}$ at the 95\% CL~\cite{BES_KsKs_2004}, respectively. In 2009 and 2012, the BESIII Collaboration searched for $J/\psi\rightarrow K^{0}_{S} K^{0}_{S}$ in 1.3 billion $J/\psi$ events and set an upper limit of $1.4\times 10^{-8}$ at the 90\% CL \cite{BESIII_KsKs_2017}.

In this paper, we search for the \textit{CP} violating decays $J/\psi\rightarrow K^{0}_{S} K^{0}_{S}$ and $\psi(3686)\rightarrow K^{0}_{S} K^{0}_{S}$ with $(10087\pm 44)\times10^{6}$ $J/\psi$ events~\cite{Ntot_Jpsi} and $(2712.4\pm 14.3)\times10^{6}$ $\psi(3686)$ events~\cite{Ntot_psip2024} accumulated by the BESIII detector. The data sample statistics are approximately 7.7 times and 190 times larger than previous studies for $J/\psi$ and $\psi(3686)$~\cite{BESIII_KsKs_2017,BES_KsKs_2004}, respectively. To avoid potential bias, a semiblind analysis is carried out using a randomly selected 10\% of the full dataset. This subset serves to evaluate the background level, verify consistency between the data and simulation samples, and validates the analysis strategy. The final results are then derived with the full dataset by applying the established analysis strategy.

\section{\boldmath BESIII Experiment and Monte Carlo Simulation}

The BESIII detector~\cite{bes3_detector} records symmetric $e^{+}e^{-}$ collisions provided by the BEPCII storage ring~\cite{BEPCII} in the center-of-mass energy range from 1.84 to 4.95~GeV, with a peak luminosity of $1.1 \times 10^{33}\;\text{cm}^{-2}\text{s}^{-1}$ achieved at $\sqrt{s} = 3.773\;\text{GeV}$.

The cylindrical core of the BESIII detector covers 93\% of the full solid angle and consists of a helium-based  multilayer drift chamber~(MDC), a time-of-flight system~(TOF), and a CsI(Tl) electromagnetic calorimeter~(EMC),
which are all enclosed in a superconducting solenoidal magnet providing a 1.0~T magnetic field. The
magnetic field was 0.9~T in 2012. The solenoid is supported by an octagonal flux-return yoke with resistive plate counter muon identification modules interleaved with steel. The charged-particle momentum resolution at $1~{\rm GeV}/c$ is $0.5\%$, and the 
${\rm d}E/{\rm d}x$ resolution is $6\%$ for electrons from Bhabha scattering. The EMC measures photon energies with a resolution of $2.5\%$ ($5\%$) at $1$~GeV in the barrel (end cap) region. The time resolution in the plastic scintillator TOF barrel region is 68~ps, while that in the end cap region is 110~ps. The end cap TOF system was upgraded in 2015 using multigap resistive plate chamber technology, providing a time resolution of 60~ps, which
benefits 87\% of the $J/\psi$ data and 83\% of the $\psi(3686)$ data used in this analysis~\cite{MRPC_TOF,*MRPC_time,*MRPC_design}.

Simulated data samples produced with a {\sc
geant4}-based~\cite{geant4,geant4_dev} Monte Carlo (MC) package, which
includes the geometric description~\cite{BesGDML, bes3_method_detector} of the BESIII detector and the detector response, are used to determine detection efficiencies and to estimate backgrounds. The simulation models the beam
energy spread and initial state radiation in the $e^+e^-$ annihilations with the generator {\sc
kkmc}~\cite{kkmc_cphc,*kkmc_prd}. 

The inclusive MC sample, consisting of $1.001\times10^{10}$ $J/\psi$ events, includes both the production of the $J/\psi$ resonance and the
continuum processes incorporated in {\sc
kkmc}. The inclusive MC sample consisting of $2.7\times 10^{9}$ $\psi(3686)$ events includes the production of the $\psi(3686)$ resonance, the initial state radiation production of the $J/\psi$, and the continuum processes incorporated in {\sc
kkmc}. All particle decays are modeled with {\sc evtgen}~\cite{evtgen_NIMA, besevtgen} using branching fractions either taken from the
Particle Data Group (PDG)~\cite{PDG2024}, when available, or otherwise estimated with {\sc lundcharm}~\cite{lundcharm, lundcharm_tuning}. Final-state radiation
from charged final-state particles is incorporated using the {\sc
photos} package~\cite{photos}. To estimate the detection efficiencies, the signal decays $J/\psi\rightarrow K^{0}_{S}K^{0}_{S}$ and  $\psi(3686)\rightarrow K^{0}_{S}K^{0}_{S}$
are generated with the VSS model, which describes the decay of a vector particle into two scalar particles~\cite{evtgen_NIMA, besevtgen}. The $K^{0}_{S}\rightarrow \pi^{+}\pi^{-}$ decays are modeled by the uniform phase space.

\section{\boldmath Event Selection}
\label{sec:selection}

Having chosen $K^{0}_{S}\rightarrow \pi^{+}\pi^{-}$, the final states of both
decays are $\pi^{+}\pi^{-}\pi^{+}\pi^{-}$. At least four charged particles with zero net charge are required to satisfy the polar angle condition $|\rm{cos}\theta|<0.93$, where $\theta$ is defined with respect to the $z$-axis, which is the
symmetry axis of the MDC. The particle identification (PID) for charged tracks is performed by combining the measurements of the specific ionization energy loss ($\mathrm{d}E/\mathrm{d}x$) in the MDC and the flight time in the TOF to form
likelihoods $L(h)$ ($h = p, K, \pi$) for each hadron $h$ hypothesis. Charged tracks are identified as pions when the pion hypothesis has the greatest likelihood.

Each $K^{0}_{S}$ candidate is reconstructed from two oppositely charged tracks assigned as $\pi^{+}$ and $\pi^{-}$. They are constrained to originate from a common vertex and are required to have an invariant mass within \mbox{$|M_{\pi^{+}\pi^{-}}-M_{K^{0}_{S}}|<18{\rm \;MeV/}c^{2}$} and $30 {\rm \;MeV/}c^{2}$ for $J/\psi\rightarrow K^{0}_{S}K^{0}_{S}$ and $\psi(3686)\rightarrow K^{0}_{S}K^{0}_{S}$, where $M_{K^{0}_{S}}$ is the $K^{0}_{S}$ nominal mass~\cite{PDG2024}. The decay length of the $K^{0}_{S}$ candidate away from the interaction point is required to be greater than twice the vertex resolution. The momenta of $\pi^{+}\pi^{-}$ are required to be within $(1.40,1.55)\;\gevc$ and $(1.70,1.84)\; \gevc$ for $J/\psi\rightarrow K^{0}_{S}K^{0}_{S}$ and  $\psi(3686)\rightarrow K^{0}_{S}K^{0}_{S}$, respectively. If there is more than one combination of two $K^{0}_{S}$ candidates, the one with the lowest sum of $\chi^{2}$ from the second vertex fit~\cite{vertexfit} is selected.

A four-constraint (4C) kinematic fit with energy and momentum conservation is performed for all possible final state combinations. To suppress the background from $\psi\rightarrow \gamma K^{0}_{S}K^{0}_{S}$, we require $\chi^{2}_{K^{0}_{S}K^{0}_{S}}<\chi^{2}_{\gamma K^{0}_{S}K^{0}_{S}}$. Events satisfying $\chi^{2}_{\rm 4C}<15$ and $\chi^{2}_{\rm 4C}<30$ are kept for $J/\psi$ and $\psi(3686)$ decays, respectively. To suppress the background from $\psi(3686) \rightarrow \pi^{+}\pi^{-}J/\psi$, \mbox{$J/\psi \rightarrow \ell^{+}\ell^{-}(\ell=e,\mu)$} due to the misidentification of leptons and pions, the mass of any combination of $\pi^{+}\pi^{-}$ pair is required to be less than 3 $\gevcc$.

The signal region is defined as 
\begin{equation}
\centering
\label{eq_5sigma}
\sqrt{(p_{K^{0}_{S1}}-p_{0})^{2}+(p_{K^{0}_{S2}}-p_{0})^{2}}<5\sigma,
\end{equation}
where the $p_{K^0_{S1}}$ and $p_{K^0_{S2}}$ are the momenta of the two $K^0_{S}$ in the $\psi$ rest frame with a  
 central value $p_{0}$ determined by fitting the corresponding signal MC samples. The $K^{0}_{S}$ momentum resolution $\sigma$ from the signal MC sample is the weighted average of the standard deviations of two Gaussian functions with the same mean. The $\sigma$ values are $1.2 \;{\rm MeV}/c$ for $J/\psi \rightarrow K^{0}_{S}K^{0}_{S}$ and $1.7 \;{\rm MeV}/c$ for $\psi(3686)\rightarrow K^{0}_{S}K^{0}_{S}$. The number of signal events is obtained by counting the remaining events within the signal region. After applying the above selection criteria, the detection efficiencies of $J/\psi \rightarrow K^{0}_{S}K^{0}_{S}$ and $\psi(3686) \rightarrow K^{0}_{S}K^{0}_{S}$ are determined to be $(23.22 \pm 0.02)\%$ and $(20.14 \pm 0.02)\%$, respectively.

\section{\boldmath Background analysis }
\label{sec:background}

The backgrounds for $\psi\rightarrow K^{0}_{S}K^{0}_{S}$ are dominated by other $\psi$ hadronic decays and the continuum process.

Studies of the $J/\psi$ inclusive MC sample show that the main backgrounds for $J/\psi\rightarrow K^{0}_{S}K^{0}_{S}$ are from $J/\psi \rightarrow \gamma K^{0}_{S} K^{0}_{S}$ and $J/\psi \rightarrow \pi^{+}\pi^{-}\pi^{+}\pi^{-}$. An exclusive MC sample of $J/\psi \rightarrow \gamma K^{0}_{S} K^{0}_{S}$ is generated using the amplitude analysis generator~\cite{pwa_gen} which provides the best agreement to data according to preliminary results from a mass-dependent analysis. Other exclusive MC samples of $J/\psi\rightarrow \pi^{+}\pi^{-}\pi^{+}\pi^{-}$, \mbox{$J/\psi\rightarrow K^{*}(892)^{0}\Bar{K^{0}}+c.c.$ } and $J/\psi\rightarrow K^{0}_{S}K^{0}_{L}$ are generated with the {\sc evtgen} models to estimate the background. Study of the $\psi(3686)$ inclusive MC sample shows that only a few background events survive after applying all selection criteria. Large exclusive MC samples of $\psi(3686)\rightarrow K^{0}_{S}K^{0}_{L}$, $\psi(3686)\rightarrow\gamma\chi_{c0,2}$($\chi_{c0,2}\rightarrow K^{0}_{S}K^{0}_{S}$), \mbox{$\psi(3686)\rightarrow\rho^{0}\pi^{+}\pi^{-}$}, $\psi(3686)\rightarrow K^{*}(892)^{0}\Bar{K^{0}}+c.c.$, $\psi(3686)\rightarrow\gamma K^{0}_{S}K^{0}_{S}$, and \mbox{$\psi(3686)\rightarrow\pi^{+}\pi^{-}J/\psi~(J/\psi\rightarrow \ell^{+}\ell^{-})$} are generated to estimate the background. According to the branching fractions from the PDG~\cite{PDG2024}, the total numbers of $J/\psi$ and $\psi(3686)$ events as well as the detection efficiencies determined with each exclusive MC sample, the total normalized background yields are $6.1 \pm 1.9$ for $J/\psi\rightarrow K^{0}_{S}K^{0}_{S}$ and  $6.7\pm 3.6$ for $\psi(3686)\rightarrow K^{0}_{S}K^{0}_{S}$.

The continuum background for $J/\psi\rightarrow K^{0}_{S}K^{0}_{S}$ is studied by using a sample of $168.3 \;\rm{pb^{-1}}$ collision data taken at $\sqrt{s}=3.08$ GeV;  the continuum contribution for $\psi(3686)\rightarrow K^{0}_{S}K^{0}_{S}$ is investigated by analyzing a data sample of $473.19 \;\rm{pb^{-1}}$  taken at $\sqrt{s}=3.65$ GeV and $2.93 \;\rm{fb^{-1}}$ of events collected at $\sqrt{s}=3.773$ GeV~\cite{lum3773}. The contribution from the continuum process is estimated with $N^{\rm{obs}}_{\rm{cont}}\times f_{c}$, where $N^{\rm{obs}}_{\rm{cont}}$ is the number of events in the 2D $K^{0}_{S}$ momentum signal region for each dataset, and $f_{c}$ denotes the scale factor that, taking into account the energy dependence of the cross section, is calculated as 
\begin{equation}    
f_{c}=\frac{\mathcal{L}_{\rm{\psi}}}{\mathcal{L}_{\rm{cont}}} \cdot  \frac{s_{\rm{cont}}}{s_{\rm{\psi}}},
\end{equation}
where $\mathcal{L}_{\rm{\psi}}$ and $s_{\rm{\psi}}$ are the corresponding integrated luminosity and the square of the center-of-mass energy for $J/\psi$ or $\psi(3686)$, the values of $\mathcal{L}_{\rm{\psi}}$ are $2962.72\;\rm{pb^{-1}}$~\cite{Ntot_Jpsi} and $3877.05\;\rm{pb^{-1}}$~\cite{Ntot_psip2024}, respectively; $\mathcal{L}_{\rm{cont}}$ and $s_{\rm{cont}}$ are the values for the data sample collected at $\sqrt{s}=3.08,\,3.65$, or $3.773$ GeV, respectively. No event survives in the signal region for any dataset.

The decays $\psi \to K^0_S K^0_L$, followed by the \textit{CP} violating decay $K^0_L \to \pi^+ \pi^-$, could pose an irreducible source of background events that cannot be distinguished from signal decays.  However, given the momenta of the $\psi$ decay daughters and the effective detector volume in which we reconstruct the neutral Kaon decay to $\pi^+ \pi^-$,  the effective decay rate of this contribution is several orders of magnitude smaller than the upper limits derived in this analysis;  these decays can therefore be safely neglected at the current level of precision.

\section{Systematic uncertainties }
 
The systematic uncertainties in the branching fraction measurements are from the total number of $\psi$ events, the quoted branching fraction, tracking of charged particles, PID, the 4C kinematic fit, and the $K^{0}_{S}$ reconstruction.

The uncertainties of the total numbers of $J/\psi$ and $\psi(3686)$ events are $0.4\%$~\cite{Ntot_Jpsi} and $0.5\%$~\cite{Ntot_psip2024}, respectively. The uncertainty of the quoted $\mathcal{B} (K^{0}_{S}\rightarrow\pi^{+}\pi^{-})$ from the PDG \cite{PDG2024} is 0.2\% for two $K^{0}_{S}$ mesons.

The control sample $J/\psi \rightarrow \pi^{+}\pi^{-}\pi^{0}$ is used to study the MDC tracking and PID uncertainties of charged pions. The difference in the tracking efficiencies between data and MC simulation is assigned as 0.5\% for each charged pion. Therefore, the systematic uncertainty due to the MDC tracking efficiency is 2.0\% for four tracks in $J/\psi$ and $\psi(3686)$ decays. The differences in two-dimensional (momentum versus polar angle) PID efficiencies between data and MC simulation of the control samples are assigned
as the systematic uncertainties for four charged pions. The resulting values are 2.7\% for $J/\psi\rightarrow K^{0}_{S}K^{0}_{S}$ and 2.5\% for $\psi(3686)\rightarrow K^{0}_{S}K^{0}_{S}$.

The control sample $\psi(3686)\rightarrow \pi^{+}\pi^{-} K^{0}_{S}K^{0}_{S}$ is used to study the correction factors for the helix parameters of charged tracks from the $K^{0}_{S}$ decays. The correction factors are determined from the pull distributions of data and MC simulation~\cite{phiksks}. The difference of signal efficiencies before and after helix correction is taken as the systematic uncertainty of the 4C kinematic fit. They are $6.2\%$ for $J/\psi\rightarrow K^{0}_{S}K^{0}_{S}$ and $1.8\%$ for $\psi(3686)\rightarrow K^{0}_{S}K^{0}_{S}$.

The systematic uncertainty from the $K^{0}_{S}$ reconstruction is estimated by using the control samples $J/\psi\rightarrow K^{0}_{S}K^{0}_{L}$ and $\psi(3686)\rightarrow K^{0}_{S}K^{0}_{L}$.  Two charged pions are selected using PID and the angle between them is required to be within $[15^{\circ},50^{\circ}]$. To improve the purity, EMC shower variables are used to distinguish signal and background with a graph neural network~\cite{Qu:2019gqs} model on the recoil side of the $\pi^{+}\pi^{-}$ system. The differences in the $K^{0}_{S}$ reconstruction efficiencies between data and MC simulation are taken as the systematic uncertainties, which are $5.2\%$ for $J/\psi\rightarrow K^{0}_{S}K^{0}_{S}$ and $7.6\%$ for $\psi(3686)\rightarrow K^{0}_{S}K^{0}_{S}$. These control samples are also used to verify the consistency between data and MC, with good agreement observed. 

\begin{table}[!htbp]
	\centering
		\setlength{\belowcaptionskip}{0.2cm}
		\caption{Relative systematic uncertainties (in \%) in the branching fraction measurements.}		
	\begin{tabular}{lcc}
		\hline \hline
  
   Source &$J/\psi\rightarrow K^{0}_{S}K^{0}_{S}$  &$\psi(3686)\rightarrow K^{0}_{S}K^{0}_{S}$  \\ \hline 
  
		$N_{\psi}$  &   0.4 &  0.5\\
		$\mathcal{B}^{\rm{PDG}}(K^{0}_{S}\rightarrow\pi^{+}\pi^{-})$    &   0.2 &   0.2   \\
		Tracking         &    2.0     & 2.0  \\
            PID                  & 2.7     & 2.5 \\
		4C kinematic fit     &  6.2   &  1.8  \\		
	    $K^{0}_{S}$ reconstruction &5.2 &7.6  \\
	
	    \hline
		{\makecell[l]{Total \\ }}    &8.8  & 8.5\\  
		\hline \hline	
	\end{tabular}

	\label{table:sysUncertainties}
\end{table}

Table~\ref{table:sysUncertainties} lists the systematic uncertainties from all sources. Assuming that each source is independent, the total systematic uncertainties are obtained by summing these uncertainties in quadrature.

\begin{figure}[htbp]
\begin{minipage}[b]{0.46\linewidth}
    \begin{overpic}[width=1.0\linewidth]{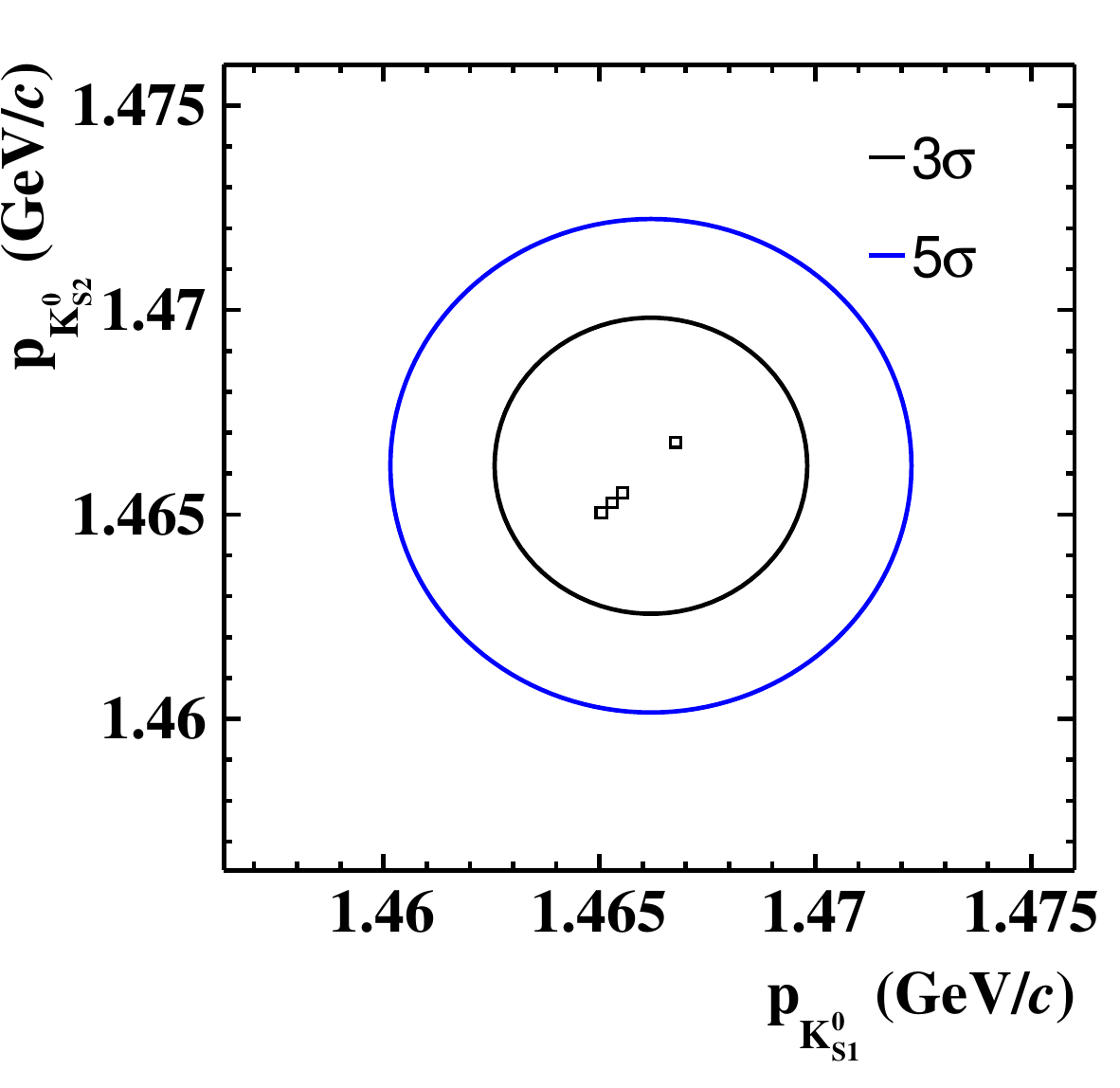}
    \put(23,80){$(a)$} 
    \end{overpic}
\end{minipage}  
\begin{minipage}[b]{0.46\linewidth}
    \begin{overpic}[width=1.0\linewidth]{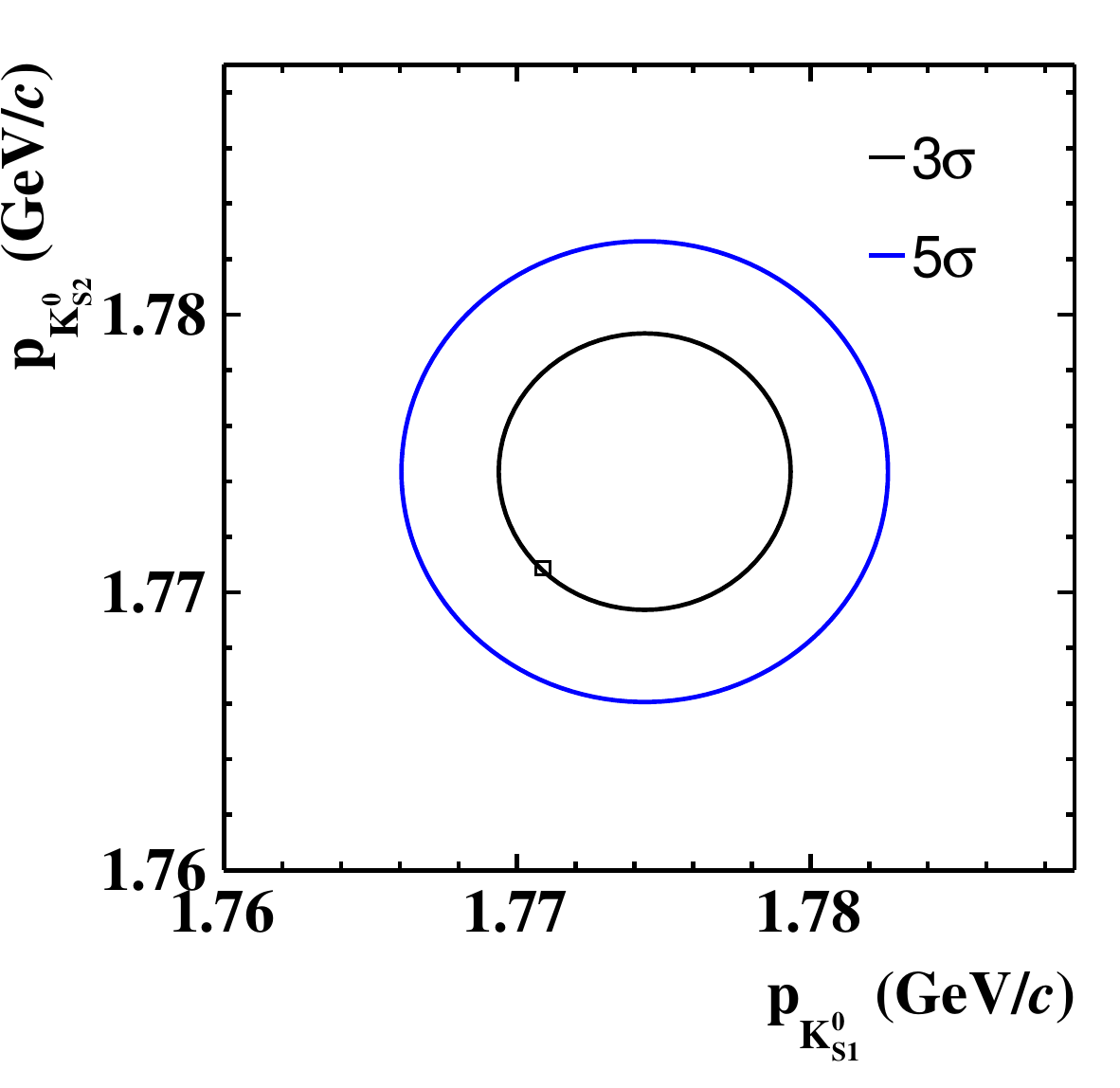}
    \put(23,80){$(b)$} 
    \end{overpic}
\end{minipage}
\caption{Two-dimensional $K^{0}_{S}$ momentum distributions for (a) $J/\psi\rightarrow K^{0}_{S}K^{0}_{S}$ and (b) $\psi(3686)\rightarrow K^{0}_{S}K^{0}_{S}$. The black and blue circles correspond to $3 \sigma$ and $5 \sigma$ determined from the individual signal MC samples, respectively.}
\label{fig:circle_data}
\end{figure}

\section{\boldmath Results}
\label{sec:results}

After unblinding all data, four candidate events for $J/\psi\rightarrow K^{0}_{S}K^{0}_{S}$ and one event for $\psi(3686)\rightarrow K^{0}_{S}K^{0}_{S}$ are observed in the corresponding signal region, as shown in Fig.~\ref{fig:circle_data}. As estimated in Sec.~\ref{sec:background}, no significant signal events are observed over the expected numbers of background events. A maximum likelihood estimator, extended from the profile likelihood approach~\cite{likelihood}, is used to determine the upper limits on the branching fractions. The likelihood function depends on the parameter of interest $\mathcal{B}(\psi\rightarrow K^{0}_{S}K^{0}_{S})$ and the nuisance parameters $\boldsymbol{\theta} =(\epsilon_{\rm{sig}},N_{\rm{bkg}})$. The likelihood function is defined as
\begin{align}
\label{eq_likelihood}
&\mathcal{L}(\mathcal{B}(\psi\rightarrow K^{0}_{S}K^{0}_{S}),\boldsymbol{\theta}) \nonumber\\
&= P(N_{\rm{obs}},\mathcal{B}(\psi\rightarrow K^{0}_{S}K^{0}_{S})\cdot N_{\psi}\cdot \mathcal{B}^{2}_{K^{0}_{S}\rightarrow \pi^{+}\pi^{-} }\cdot \epsilon_{\rm{sig}} \nonumber\\ & + \sum_{i} N_{{\rm{bkg}},i})
\cdot \prod_{i} P(N^{\rm{exp}}_{{\rm{bkg}},i},{N}_{{\rm{bkg}},i})\cdot G(\epsilon^{\rm{MC}}_{\rm{sig}},{\epsilon}_{\rm{sig}},\sigma_{\epsilon^{\rm{MC}}_{\rm{sig}}}),
\end{align} 

where the number of observed events $N_{\rm obs}$ is assumed to follow a Poisson distribution $(P)$. Here, the detection efficiency $\epsilon_{\rm{sig}}$ follows a Gaussian distribution $(G)$ with mean value $\epsilon^{\rm{MC}}_{\rm{sig}}$ and uncertainty $\sigma_{\epsilon^{\rm{MC}}_{\rm{sig}}}$, which are determined from signal MC samples and systematic uncertainties studies. The number of background events $N_{\rm{bkg}}$ obeys a Poisson distribution $(P)$ with expected value $N^{\rm{exp}}_{\rm{bkg}}$, which is determined from the background study. The subscript $i$ represents all background types, including those from $\psi$ decay and continuum. The number of $\psi$ events $N_{\psi}$, is obtained from an analysis of the inclusive $J/\psi$ [$\psi(3686)$] sample~\cite{Ntot_Jpsi,Ntot_psip2024}.

\begin{figure}[htbp!]
\centering
\begin{minipage}[b]{1.0\linewidth}
\vspace{3pt}
    \begin{overpic}[width=0.8\linewidth]{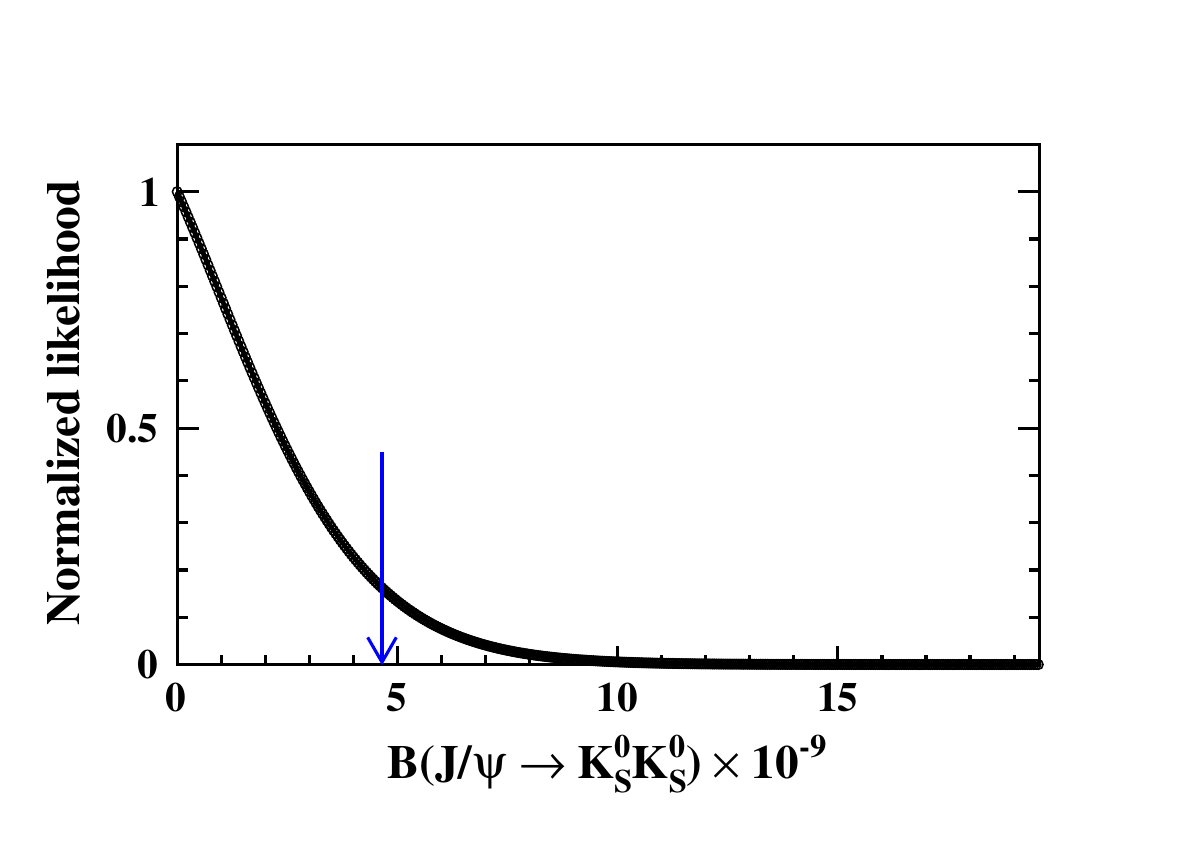}
    \put(75,54){$(a)$} 
    \end{overpic}
\end{minipage}  
\begin{minipage}[b]{1.0\linewidth}
\vspace{-10pt}
    \begin{overpic}[width=0.8\linewidth]{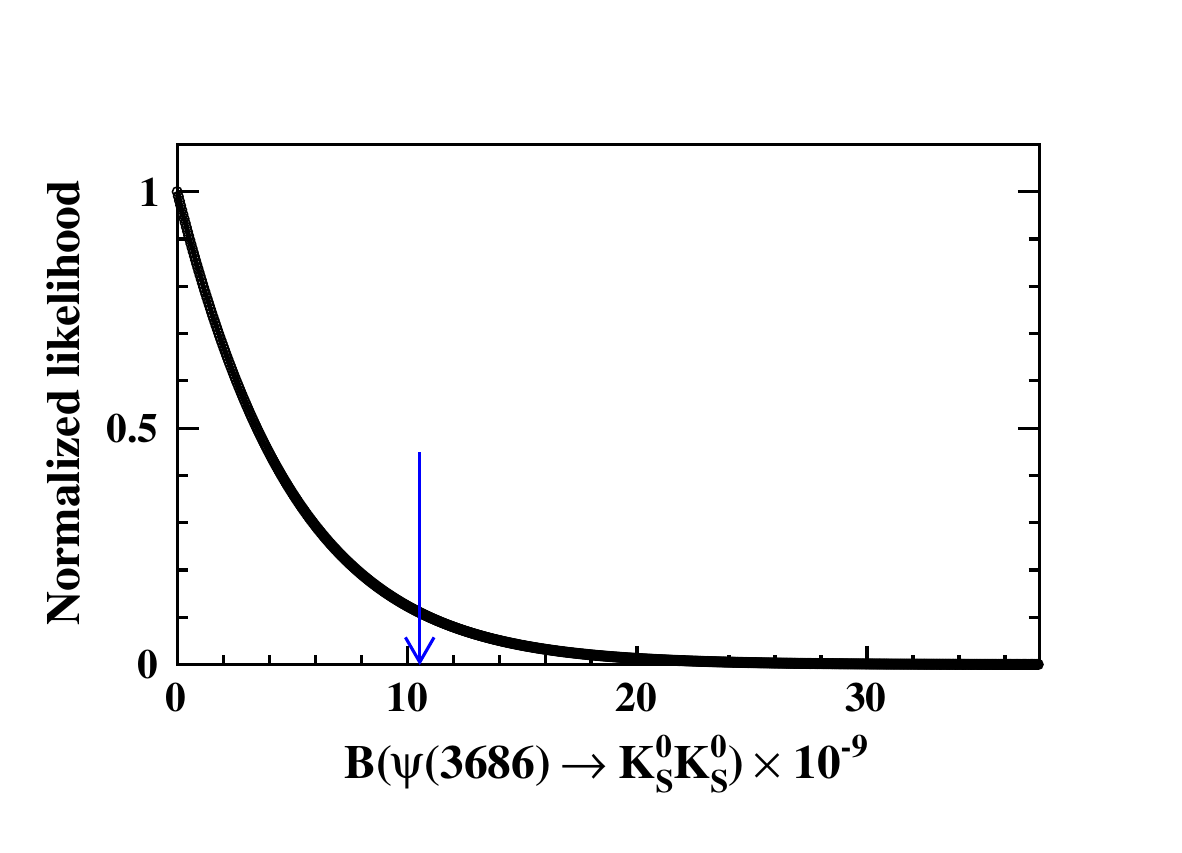}
    \put(75,54){$(b)$} 
    \end{overpic}
\end{minipage}
\caption{The normalized likelihood distributions for (a) $J/\psi\rightarrow K^{0}_{S}K^{0}_{S}$ and (b) $\psi(3686)\rightarrow K^{0}_{S}K^{0}_{S}$. The blue arrows indicate the 90\% CL upper limits.}
\label{fig:upl_data}
\end{figure}

Fig.~\ref{fig:upl_data} shows the likelihood curves for $J/\psi\rightarrow K^{0}_{S}K^{0}_{S}$ and $\psi(3686)\rightarrow K^{0}_{S}K^{0}_{S}$. We implicitly use a flat prior for the branching fraction in the determination of the upper limits.
After integrating the likelihood distributions with all factors included, the upper limits of the branching fractions at the 90\% CL after considering systematic uncertainties are determined to be $\mathcal{B}(J/\psi\rightarrow K^{0}_{S}K^{0}_{S}) < 4.7 \times 10^{-9}$ and $\mathcal{B}(\psi(3686)\rightarrow K^{0}_{S}K^{0}_{S}) < 1.1 \times 10^{-8}$ as listed in Table \ref{table:upLimit_full}.

\begin{table}[htbp]
	\centering
	\setlength{\belowcaptionskip}{0.2cm}
		\caption{The observed yields in data ($N_{\rm obs}$), estimated background yield ($N^{\rm{exp}}_{\rm{bkg}}$), detection efficiencies ($\epsilon^{\rm{MC}}_{\rm{sig}}$) and their uncertainties ($\sigma_{\epsilon^{\rm{MC}}_{\rm{sig}}}$) including both statistical and systematic uncertainties,  and the upper limits on branching fractions of $J/\psi\rightarrow K^{0}_{S}K^{0}_{S}$ and $\psi(3686)\rightarrow K^{0}_{S}K^{0}_{S}$. All upper limits are set at the 90\% CL.} 	
	\begin{tabular}{l c c}
		\hline \hline
		Channel & $J/\psi\rightarrow K^{0}_{S}K^{0}_{S}$ & $\psi(3686)\rightarrow K^{0}_{S}K^{0}_{S}$    \\
		\hline 
		$N_{\rm{obs}}$ &  4  & 1     \\
		$N^{\rm{exp}}_{\rm{bkg}}$   & $6.1 \pm  1.9$  & $6.7 \pm 3.6$    \\
		$\epsilon^{\rm{MC}}_{\rm{sig}}$    &  23.22\% & 20.14\% \\
		$\sigma_{\epsilon^{\rm{MC}}_{\rm{sig}}}$    & 2.04\%  & 1.71\%   \\	    

	    $\mathcal{B_{\rm UL}}$    &  $4.7\times10^{-9}$  & $1.1\times10^{-8}$  \\	    
		\hline \hline	
	\end{tabular}
	\label{table:upLimit_full}
\end{table}

\section{\boldmath Summary}
\label{sec:summary}

Using $(10087\pm 44)\times10^{6}$ $J/\psi$ events and $(2712.4\pm 14.3)\times10^{6}$ $\psi(3686)$ events collected by the BESIII detector,  we have searched for the \textit{CP} violating decays $J/\psi\rightarrow K^{0}_{S}K^{0}_{S}$ and $\psi(3686)\rightarrow K^{0}_{S}K^{0}_{S}$. A semiblind analysis finds no significant excess in the datasets with respect to the expected background yields. The upper limits on their branching fractions at the 90\% CL are determined to be $\mathcal{B}(J/\psi\rightarrow K^{0}_{S}K^{0}_{S}) < 4.7 \times 10^{-9}$ and $\mathcal{B}(\psi(3686)\rightarrow K^{0}_{S}K^{0}_{S}) < 1.1 \times 10^{-8}$, where the systematic uncertainties have been taken into account. The results improve the previous best limits~\cite{BESIII_KsKs_2017,BES_KsKs_2004} by a factor of three for $J/\psi\rightarrow K^{0}_{S}K^{0}_{S}$ and two orders of magnitude for $\psi(3686)\rightarrow K^{0}_{S}K^{0}_{S}$, which can be used to constrain new physics parameters. Experimental results have already excluded the EPR locality expectations~\cite{EPR_KsKs_2021}. These upper limits reach the level of the \textit{CP} violation expectations~\cite{CPV_psip_2006} and could provide experimental evidence for further theoretical studies.  Using the ratio of the branching fractions of $J/\psi$ and $\psi(3686)$ into $K^{0}_{S} K^{0}_{S}$ and $K^{0}_{S} K^{0}_{L}$~\cite{PDG2024}, the \textit{CPT} violation parameter $|\omega|$ is estimated to be $(4.91\pm0.14)\times10^{-3}$ and $(1.44\pm0.04)\times10^{-2}$ for the respective upper limits of $J/\psi\rightarrow K^{0}_{S}K^{0}_{S}$ and $\psi(3686)\rightarrow K^{0}_{S}K^{0}_{S}$.

\acknowledgments

The BESIII Collaboration thanks the staff of BEPCII and the IHEP computing center for their strong support. This work is supported in part by National Key R\&D Program of China under Contracts No. 2023YFA1606000; Joint Large-Scale Scientific Facility Funds of the NSFC and CAS under Contract No. U1832207; National Natural Science Foundation of China (NSFC) under Contracts No. 11635010, No. 11735014, No. 11935015, No. 11935016, No. 11935018, No. 12025502, No. 12035009, No. 12035013, No. 12061131003, No. 12192260, No. 12192261, No. 12192262, No. 12192263, No. 12192264, No. 12192265, No. 12221005, No. 12225509, No. 12235017, No. 12361141819; the Chinese Academy of Sciences (CAS) Large-Scale Scientific Facility Program; the CAS Center for Excellence in Particle Physics (CCEPP); CAS under Contract No. YSBR-101; 100 Talents Program of CAS; The Institute of Nuclear and Particle Physics (INPAC) and Shanghai Key Laboratory for Particle Physics and Cosmology; Agencia Nacional de Investigación y Desarrollo de Chile (ANID), Chile under Contract No. ANID PIA/APOYO AFB230003; German Research Foundation DFG under Contract No. FOR5327; Istituto Nazionale di Fisica Nucleare, Italy; Knut and Alice Wallenberg Foundation under Contracts No. 2021.0174, No. 2021.0299; Ministry of Development of Turkey under Contract No. DPT2006K-120470; National Research Foundation of Korea under Contract No. NRF-2022R1A2C1092335; National Science and Technology fund of Mongolia; National Science Research and Innovation Fund (NSRF) via the Program Management Unit for Human Resources \& Institutional Development, Research and Innovation of Thailand under Contract No. B50G670107; Polish National Science Centre under Contract No. 2019/35/O/ST2/02907; Swedish Research Council under Contract No. 2019.04595; The Swedish Foundation for International Cooperation in Research and Higher Education under Contract No. CH2018-7756; U. S. Department of Energy under Contract No. DE-FG02-05ER41374

\bibliographystyle{apsrev4-2}
\bibliography{mybib}
\end{document}